\def\gta{\mathrel{\raise 0.1em \hbox{$>$} \hskip -0.8em \lower 0.4em
   \hbox{$\sim$}}}
\def\lta{\mathrel{\vcenter{\vbox{\offinterlineskip \hbox{$<$}
     \vskip 0.2 pt \hbox{$\sim$}}}}}
\def\tot{{\rm tot}}

\documentclass[12pt]{iopart}
\usepackage{epsf}
\usepackage{hyperref}
\usepackage[square,comma,sort&compress]{natbib}
\usepackage{hypernat}         

\begin{document}

\rightline{MIT-CTP\#3811 \qquad hep-th/0702178}

\title[Eternal inflation and its implications]{Eternal inflation
and its implications\footnote{Talk presented at the {\it 2nd
International Conference on Quantum Theories and Renormalization
Group in Gravity and Cosmology} {\bf (IRGAC 2006),} Barcelona, Spain,
11-15 July 2006, to be published in {\it J.~Phys.~A.}}}

\author{Alan H. Guth}

\address{Center for Theoretical Physics, Laboratory for
Nuclear Science, and Department of Physics, Massachusetts
Institute of Technology, Cambridge, MA 02139}
\ead{guth@ctp.mit.edu}
\begin{abstract}
I summarize the arguments that strongly suggest that our
universe is the product of inflation.  The mechanisms that lead
to eternal inflation in both new and chaotic models are
described.  Although the infinity of pocket universes produced by
eternal inflation are unobservable, it is argued that eternal
inflation has real consequences in terms of the way that
predictions are extracted from theoretical models.  The
ambiguities in defining probabilities in eternally inflating
spacetimes are reviewed, with emphasis on the youngness paradox
that results from a synchronous gauge regularization technique. 
Although inflation is generically eternal into the future, it is
not eternal into the past: it can be proven under reasonable
assumptions that the inflating region must be incomplete in past
directions, so some physics other than inflation is needed to
describe the past boundary of the inflating region. 

\end{abstract}

\maketitle

\section{Introduction: the successes of inflation}

Since the proposal of the inflationary model some 25 years ago
\cite{Guth1, Linde1, Albrecht-Steinhardt1, Starobinsky1}, 
inflation has been remarkably successful in explaining many
important qualitative and quantitative properties of the
universe.  In this article I will summarize the key successes,
and then discuss a number of issues associated with the eternal
nature of inflation.  In my opinion, the evidence that our
universe is the result of some form of inflation is very solid. 
Since the term {\it inflation} encompasses a wide range of
detailed theories, it is hard to imagine any reasonable
alternative.  The basic arguments are as follows:

\begin{enumerate}
\item{\it The universe is big}

\qquad First of all, we know that the universe is incredibly
large: the visible part of the universe contains about $10^{90}$
particles.  Since we have all grown up in a large universe, it is
easy to take this fact for granted: of course the universe is
big, it's the whole universe! In ``standard'' FRW cosmology,
without inflation, one simply postulates that about $10^{90}$ or
more particles were here from the start.  However, in the context
of present-day cosmology, many of us hope that even the creation
of the universe can be described in scientific terms.  Thus, we
are led to at least think about a theory that might explain how
the universe got to be so big.  Whatever that theory is, it has
to somehow explain the number of particles, $10^{90}$ or more. 
However, it is hard to imagine such a number arising from a
calculation in which the input consists only of geometrical
quantities, quantities associated with simple dynamics, and
factors of 2 or $\pi$.  The easiest way by far to get a huge
number, with only modest numbers as input, is for the calculation
to involve an exponential.  The exponential expansion of
inflation reduces the problem of explaining $10^{90}$ particles
to the problem of explaining 60 or 70 e-foldings of inflation. 
In fact, it is easy to construct underlying particle theories
that will give far more than 70 e-foldings of inflation. 
Inflationary cosmology therefore suggests that, even though the
observed universe is incredibly large, it is only an
infinitesimal fraction of the entire universe.

\item{\it The Hubble expansion}

\qquad The Hubble expansion is also easy to take for granted,
since we have all known about it from our earliest readings in
cosmology.  In standard FRW cosmology, the Hubble expansion is
part of the list of postulates that define the initial
conditions.  But inflation actually offers the possibility of
explaining how the Hubble expansion began.  The repulsive gravity
associated with the false vacuum is just what Hubble ordered.  It
is exactly the kind of force needed to propel the universe into a
pattern of motion in which any two particles are moving apart
with a velocity proportional to their separation.

\item{\it Homogeneity and isotropy}

\qquad The degree of uniformity in the universe is startling. 
The intensity of the cosmic background radiation is the same in
all directions, after it is corrected for the motion of the
Earth, to the incredible precision of one part in 100,000.  To
get some feeling for how high this precision is, we can imagine a
marble that is spherical to one part in 100,000.  The surface of
the marble would have to be shaped to an accuracy of about 1,000
angstroms, a quarter of the wavelength of light. 

\qquad Although modern technology makes it possible to grind
lenses to quarter-wavelength accuracy, we would nonetheless be
shocked if we unearthed a stone, produced by natural processes,
that was round to an accuracy of 1,000 angstroms.  If we try to
imagine that such a stone were found, I am sure that no one would
accept an explanation of its origin which simply proposed that
the stone started out perfectly round.  Similarly, I do not think
it makes sense to consider any theory of cosmogenesis that cannot
offer some explanation of how the universe became so incredibly
isotropic. 

\qquad The cosmic background radiation was released about 300,000
years after the big bang, after the universe cooled enough so
that the opaque plasma neutralized into a transparent gas.  The
cosmic background radiation photons have mostly been traveling on
straight lines since then, so they provide an image of what the
universe looked like at 300,000 years after the big bang.  The
observed uniformity of the radiation therefore implies that the
observed universe had become uniform in temperature by that time. 
In standard FRW cosmology, a simple calculation shows that the
uniformity could be established so quickly only if signals could
propagate at 100 times the speed of light, a proposition clearly
contradicting the known laws of physics.  In inflationary
cosmology, however, the uniformity is easily explained.  The
uniformity is created initially on microscopic scales, by normal
thermal-equilibrium processes, and then inflation takes over and
stretches the regions of uniformity to become large enough to
encompass the observed universe.

\item{\it The flatness problem}

\qquad I find the flatness problem particularly impressive,
because of the extraordinary numbers that it involves. The
problem concerns the value of the ratio
\begin{equation}
  \Omega_\tot \equiv {\rho_\tot \over \rho_c} \ ,
  \label{eq:13}
\end{equation}
where $\rho_\tot$ is the average total mass density of the
universe and $\rho_c = 3 H^2 / 8 \pi G$ is the critical density,
the density that would make the universe spatially flat.  (In the
definition of ``total mass density,'' I am including the vacuum
energy $\rho_{\rm vac} = \Lambda/ 8 \pi G$ associated with the
cosmological constant $\Lambda$, if it is nonzero.)

\qquad By combining data from the Wilkinson Microwave Anisotropy
Probe (WMAP), the Sloan Digital Sky Survey (SDSS), and
observations of type Ia supernovae, the authors of
Ref.~\cite{Tegmark2004} deduced that the present value of
$\Omega_\tot$ is equal to one within a few percent ($\Omega_\tot =
1.012^{+0.018}_{-0.022}$).  Although this value is very close to
one, the really stringent constraint comes from extrapolating
$\Omega_\tot$ to early times, since $\Omega_\tot=1$ is an unstable
equilibrium point of the standard model evolution.  Thus, if
$\Omega_\tot$ was ever {\it exactly} equal to one, it would remain
exactly one forever.  However, if $\Omega_\tot$ differed slightly from
one in the early universe, that difference---whether positive or
negative---would be amplified with time.  In particular, it can
be shown that $\Omega_\tot - 1$ grows as
\begin{equation}
  \Omega_\tot - 1 \propto \cases{t &(during the radiation-dominated era)\cr
    t^{2/3} &(during the matter-dominated era)\ .\cr}
  \label{eq:15}
\end{equation}
Dicke and Peebles \cite{dicke} pointed out that at $t=1$ second,
for example, when the processes of big bang nucleosynthesis were
just beginning, $\Omega_\tot$ must have equaled one to an
accuracy of one part in $10^{15}$.  Classical cosmology provides
no explanation for this fact---it is simply assumed as part of
the initial conditions.  In the context of modern particle
theory, where we try to push things all the way back to the
Planck time, $10^{-43}$ s, the problem becomes even more extreme. 
If one specifies the value of $\Omega_\tot$ at the Planck time,
it has to equal one to 59 decimal places in order to be in the
allowed range today. 

\qquad While this extraordinary flatness of the early universe
has no explanation in classical FRW cosmology, it is a natural
prediction for inflationary cosmology.  During the inflationary
period, instead of $\Omega_\tot$ being driven away from one as
described by Eq.~(\ref{eq:15}), $\Omega_\tot$ is driven towards one,
with exponential swiftness:
\begin{equation}
  \Omega_\tot - 1 \propto e^{-2 H_{\rm inf} t} \ ,
  \label{eq:16}
\end{equation}
where $H_{\rm inf}$ is the Hubble parameter during inflation. 
Thus, as long as there is a long enough period of inflation,
$\Omega_\tot$ can start at almost any value, and it will be driven to
unity by the exponential expansion. 

\item{\it Absence of magnetic monopoles}

\qquad All grand unified theories predict that there should be,
in the spectrum of possible particles, extremely massive
particles carrying a net magnetic charge.  By combining grand
unified theories with classical cosmology without inflation,
Preskill \cite{preskill} found that magnetic monopoles would be
produced so copiously that they would outweigh everything else in
the universe by a factor of about $10^{12}$.  A mass density this
large would cause the inferred age of the universe to drop to
about 30,000 years! Inflation is certainly the simplest known
mechanism to eliminate monopoles from the visible universe, even
though they are still in the spectrum of possible particles.  The
monopoles are eliminated simply by arranging the parameters so
that inflation takes place after (or during) monopole production,
so the monopole density is diluted to a completely negligible
level.

\begin{figure}
\centerline{\epsfbox{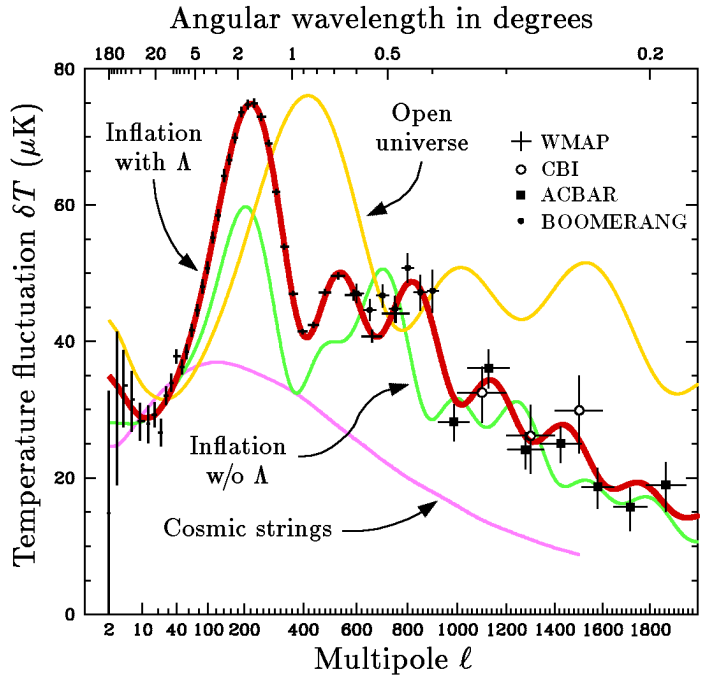}}
\caption{Comparison of the latest observational measurements of
the temperature fluctuations in the CMB with several theoretical
models, as described in the text.  The temperature pattern on the
sky is expanded in multipoles (i.e., spherical harmonics), and
the intensity is plotted as a function of the multipole number
$\ell$.  Roughly speaking, each multipole $\ell$ corresponds to
ripples with an angular wavelength of $360^\circ/\ell$.} 
\label{cmb12}
\end{figure}

\item{\it Anisotropy of the cosmic background radiation}

\qquad The process of inflation smooths the universe essentially
completely, but density fluctuations are generated as inflation
ends by the quantum fluctuations of the inflaton field
\cite{FluctuationHistory, FluctuationReviews}. Generically these
are adiabatic Gaussian fluctuations with a nearly scale-invariant
spectrum.

\qquad Until recently, astronomers were aware of several
cosmological models that were consistent with the known data: an
open universe, with $\Omega \cong 0.3$; an inflationary universe
with considerable dark energy ($\Lambda$); an inflationary
universe without $\Lambda$; and a universe in which the
primordial perturbations arose from topological defects such as
cosmic strings.  Each of these models leads to a distinctive
pattern of resonant oscillations in the early universe, which can
be probed today through its imprint on the CMB\null.  As can be
seen in Fig.~\ref{cmb12} \cite{FigCMBinfo}, three of the models are
now definitively ruled out.  The full class of inflationary
models can make a variety of predictions, but the predictions of
the simplest inflationary models with large $\Lambda$, shown on
the graph, fit the data beautifully.

\end{enumerate}

\section{Eternal Inflation: Mechanisms}

The remainder of this article will discuss eternal
inflation---the questions that it can answer, and the questions
that it raises.  In this section I discuss the mechanisms that
make eternal inflation possible, leaving the other issues for the
following sections.  I will discuss eternal inflation first in
the context of new inflation, and then in the context of chaotic
inflation, where it is more subtle. 

\begin{figure}
\epsfxsize=190pt
\centerline{\epsfbox{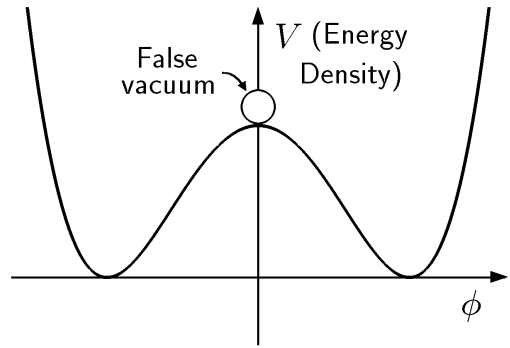}}
\caption{Evolution of the inflaton field during new
inflation.}
\label{newpotential}
\end{figure}

\subsection{Eternal New Inflation}

The eternal nature of new inflation was first discovered by
Steinhardt \cite{steinhardt-nuffield}, and later that year Vilenkin
\cite{vilenkin-eternal} showed that new inflationary models are
generically eternal.  Although the false vacuum is a metastable
state, the decay of the false vacuum is an exponential process,
very much like the decay of any radioactive or unstable
substance.  The probability of finding the inflaton field at the
top of the plateau in its potential energy diagram,
Fig.~\ref{newpotential}, does not fall sharply to zero, but
instead trails off exponentially with time \cite{guth-pi2}. 
However, unlike a normal radioactive substance, the false vacuum
exponentially expands at the same time that it decays. In fact,
in any successful inflationary model the rate of exponential
expansion is always much faster than the rate of exponential
decay.  Therefore, even though the false vacuum is decaying, it
never disappears, and in fact the total volume of the false
vacuum, once inflation starts, continues to grow exponentially
with time, ad infinitum. 

\begin{figure}
\centerline{\epsfbox{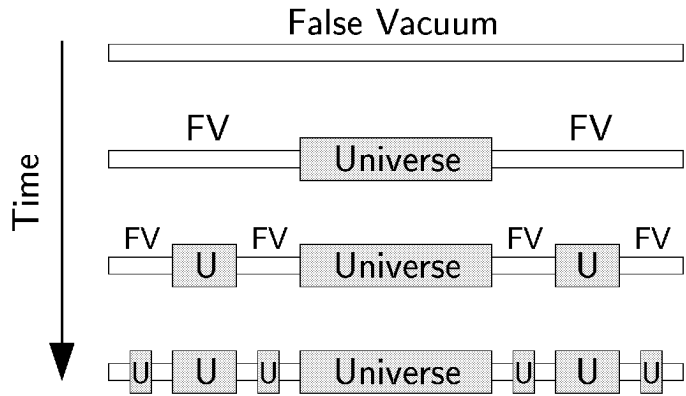}}
\caption{A schematic illustration of eternal inflation.} 
\label{eternalline}
\end{figure}

Fig.~\ref{eternalline} shows a schematic diagram of an eternally
inflating universe.  The top bar indicates a region of false
vacuum.  The evolution of this region is shown by the successive
bars moving downward, except that the expansion could not be
shown while still fitting all the bars on the page.  So the
region is shown as having a fixed size in comoving coordinates,
while the scale factor, which is not shown, increases from each
bar to the next.  As a concrete example, suppose that the scale
factor for each bar is three times larger than for the previous
bar.  If we follow the region of false vacuum as it evolves from
the situation shown in the top bar to the situation shown in the
second bar, in about one third of the region the scalar field
rolls down the hill of the potential energy diagram,
precipitating a local big bang that will evolve into something
that will eventually appear to its inhabitants as a universe. 
This local big bang region is shown in gray and labelled
``Universe.'' Meanwhile, however, the space has expanded so much
that each of the two remaining regions of false vacuum is the
same size as the starting region.  Thus, if we follow the region
for another time interval of the same duration, each of these
regions of false vacuum will break up, with about one third of
each evolving into a local universe, as shown on the third bar
from the top.  Now there are four remaining regions of false
vacuum, and again each is as large as the starting region.  This
process will repeat itself literally forever, producing a kind of
a fractal structure to the universe, resulting in an infinite
number of the local universes shown in gray.  There is no
standard name for these local universes, but they are often
called bubble universes.  I prefer, however, to call them pocket
universes, to avoid the suggestion that they are round.  While
bubbles formed in first-order phase transitions are round
\cite{coleman-deluccia}, the local universes formed in eternal
new inflation are generally very irregular, as can be seen for
example in the two-dimensional simulation by Vanchurin, Vilenkin,
and Winitzki in Fig.~2 of Ref.~\cite{vvw}.

The diagram in Fig.~\ref{eternalline} is of course an
idealization.  The real universe is three dimensional, while the
diagram illustrates a schematic one-dimensional universe.  It is
also important that the decay of the false vacuum is really a
random process, while the diagram was constructed to show a very
systematic decay, because it is easier to draw and to think
about.  When these inaccuracies are corrected, we are still left
with a scenario in which inflation leads asymptotically to a
fractal structure \cite{aryal-vilenkin} in which the universe as
a whole is populated by pocket universes on arbitrarily small
comoving scales.  Of course this fractal structure is entirely on
distance scales much too large to be observed, so we cannot
expect astronomers to see it.  Nonetheless, one does have to
think about the fractal structure if one wants to understand the
very large scale structure of the spacetime produced by
inflation. 

Most important of all is the simple statement that once inflation
happens, it produces not just one universe, but an infinite
number of universes. 

\begin{figure}
\epsfxsize=260pt
\centerline{\epsfbox{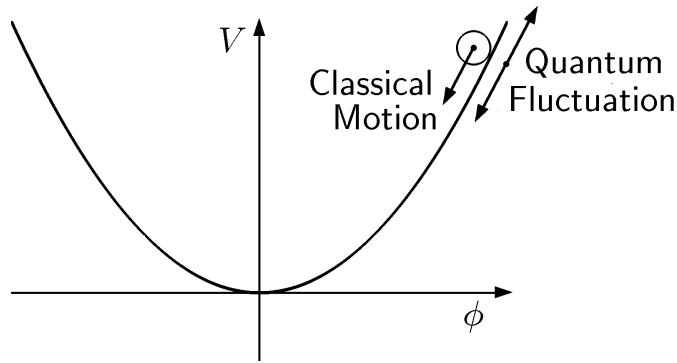}}
\caption{Evolution of the inflaton field during eternal chaotic
inflation.}
\label{chaotic-eternal}
\end{figure}

\subsection{Eternal Chaotic Inflation:}

The eternal nature of new inflation depends crucially on the
scalar field lingering at the top of the plateau of
Fig.~\ref{newpotential}.  Since the potential function for
chaotic inflation, Fig.~\ref{chaotic-eternal}, does not have a
plateau, it is not obvious how eternal inflation can happen in
this context.  Nonetheless, Andrei Linde \cite{linde-eternal}
showed in 1986 that chaotic inflation can also be eternal. 

In this case inflation occurs as the scalar field rolls down a
hill of the potential energy diagram, as in
Fig.~\ref{chaotic-eternal}, starting high on the hill.  As the field
rolls down the hill, quantum fluctuations will be superimposed on
top of the classical motion.  The best way to think about this is
to ask what happens during one time interval of duration $\Delta
t = H^{-1}$ (one Hubble time), in a region of one Hubble volume
$H^{-3}$.  Suppose that $\phi_0$ is the average value of $\phi$
in this region, at the start of the time interval.  By the
definition of a Hubble time, we know how much expansion is going
to occur during the time interval: exactly a factor of $e$. 
(This is the only exact number in today's talk, so I wanted to
emphasize the point.)  That means the volume will expand by a
factor of $e^3$.  One of the deep truths that one learns by
working on inflation is that $e^3$ is about equal to 20, so the
volume will expand by a factor of 20.  Since correlations
typically extend over about a Hubble length, by the end of one
Hubble time, the initial Hubble-sized region grows and breaks up
into 20 independent Hubble-sized regions. 

As the scalar field is classically rolling down the hill, the
change in the field $\Delta \phi$ during the time interval
$\Delta t$ is going to be modified by quantum fluctuations
$\Delta \phi_{\rm qu}$, which can drive the field upward or
downward relative to the classical trajectory.  For any one of
the 20 regions at the end of the time interval, we can describe
the change in $\phi$ during the interval by
\begin{equation}
  \Delta \phi = \Delta \phi_{\rm cl} + \Delta \phi_{\rm qu} \ ,
\end{equation}
where $\Delta \phi_{\rm cl}$ is the classical value of $\Delta
\phi$. In lowest order perturbation theory the fluctuations are
calculated using free quantum field, which implies that $\Delta
\phi_{\rm qu}$, the quantum fluctuation averaged over one of the
20 Hubble volumes at the end, will have a Gaussian probability
distribution, with a width of order $H/2 \pi$
\cite{random-vil-ford,random-linde,Starobinsky2,random-starobinsky}.
There is then always some probability that the sum of the two
terms on the right-hand side will be positive --- that the scalar
field will fluctuate up and not down.  As long as that
probability is bigger than 1 in 20, then the number of inflating
regions with $\phi \ge \phi_0$ will be larger at the end of the
time interval $\Delta t$ than it was at the beginning.  This
process will then go on forever, so inflation will never end. 

Thus, the criterion for eternal inflation is that the probability
for the scalar field to go up must be bigger than $1/e^3 \approx
1/20$.  For a Gaussian probability distribution, this condition
will be met provided that the standard deviation for $\Delta
\phi_{\rm qu} $ is bigger than $0.61 |\Delta \phi_{\rm cl}|$.
Using $\Delta \phi_{\rm cl} \approx \dot \phi_{\rm cl} H^{-1}$,
the criterion becomes
\begin{equation}
  \Delta \phi_{\rm qu} \approx {H \over 2 \pi} > 0.61 \, | \dot
     \phi_{\rm cl}| \, H^{-1} \Longleftrightarrow {H^2 \over |\dot
     \phi_{\rm cl}|} > 3.8 \ . 
 \label{eq:33}
\end{equation}
We have not discussed the calculation of density perturbations in
detail, but the condition (\ref{eq:33}) for eternal inflation is
equivalent to the condition that $\delta \rho/\rho$ on ultra-long
length scales is bigger than a number of order unity.
 
The probability that $\Delta \phi$ is positive tends to increase
as one considers larger and larger values of $\phi$, so sooner or
later one reaches the point at which inflation becomes eternal. 
If one takes, for example, a scalar field with a potential
\begin{equation}
  V(\phi) = {1 \over 4} \lambda \phi^4 \ ,
  \label{eq:34}
\end{equation}
then the de Sitter space equation of motion in flat
Robertson-Walker coordinates takes the form
\begin{equation}
  \ddot \phi + 3 H \dot \phi = - \lambda \phi^3 \ ,
  \label{eq:35}
\end{equation}
where spatial derivatives have been neglected.  In the
``slow-roll'' approximation one also neglects the $\ddot \phi$
term, so $\dot \phi \approx - \lambda \phi^3 / (3 H)$, where the
Hubble constant $H$ is related to the energy density by
\begin{equation}
  H^2 = {8 \pi \over 3} G \rho = {2 \pi \over 3} {\lambda \phi^4
     \over M_p^2} \ .
  \label{eq:36}
\end{equation}
Putting these relations together, one finds that the criterion
for eternal inflation, Eq.~(\ref{eq:33}), becomes
\begin{equation}
  \phi > 0.75 \, \lambda^{-1/6} \, M_p \ . 
 \label{eq:37}
\end{equation}

Since $\lambda$ must be taken very small, on the order of
$10^{-12}$, for the density perturbations to have the right
magnitude, this value for the field is generally well above the
Planck scale.  The corresponding energy density, however, is
given by
\begin{equation}
  V(\phi) = {1 \over 4} \lambda \phi^4 = .079 \lambda^{1/3} M_p^4
     \ ,
  \label{eq:38}
\end{equation}
which is actually far below the Planck scale.

So for these reasons we think inflation is almost always eternal. 
I think the inevitability of eternal inflation in the context of
new inflation is really unassailable --- I do not see how it
could possibly be avoided, assuming that the rolling of the
scalar field off the top of the hill is slow enough to allow
inflation to be successful.  The argument in the case of chaotic
inflation is less rigorous, but I still feel confident that it is
essentially correct.  For eternal inflation to set in, all one
needs is that the probability for the field to increase in a
given Hubble-sized volume during a Hubble time interval is larger
than 1/20.

Thus, once inflation happens, it produces not just one universe,
but an infinite number of universes. 

\section{Implications for the Landscape of String Theory}

Until recently, the idea of eternal inflation was viewed by most
physicists as an oddity, of interest only to a small subset of
cosmologists who were afraid to deal with concepts that make real
contact with observation.  The role of eternal inflation in
scientific thinking, however, was greatly boosted by the
realization that string theory has no preferred vacuum, but
instead has perhaps $10^{1000}$ \cite{Bousso-Polchinksi,
Susskind-Landscape} metastable vacuum-like states.  Eternal
inflation then has potentially a direct impact on fundamental
physics, since it can provide a mechanism to populate the
landscape of string vacua.  While all of these vacua are
described by the same fundamental string theory, the apparent
laws of physics at low energies could differ dramatically from
one vacuum to another.  In particular, the value of the
cosmological constant (e.g., the vacuum energy density) would be
expected to have different values for different vacua.

The combination of the string landscape with eternal inflation
has in turn led to a markedly increased interest in anthropic
reasoning, since we now have a respectable set of theoretical
ideas that provide a setting for such reasoning.  To many
physicists, the new setting for anthropic reasoning is a welcome
opportunity: in the multiverse, life will evolve only in very
rare regions where the local laws of physics just happen to have
the properties needed for life, giving a simple explanation for
why the observed universe appears to have just the right
properties for the evolution of life.  The incredibly small value
of the cosmological constant is a telling example of a feature
that seems to be needed for life, but for which an explanation
from fundamental physics is painfully lacking.  Anthropic
reasoning can give the illusion of intelligent design
\cite{Susskind-book}, without the need for any intelligent
intervention. 

On the other hand, many other physicists have an abhorrence of
anthropic reasoning.  To this group, anthropic reasoning means
the end of the hope that precise and unique predictions can be
made on the basis of logical deduction \cite{Gross-reviews}. 
Since this hope should not be given up lightly, many physicists
are still trying to find some mechanism to pick out a unique
vacuum from string theory.  So far there is no discernable
progress.

It seems sensible, to me, to consider anthropic reasoning to be
the explanation of last resort.  That is, in the absence of any
detailed understanding of the multiverse, life, or the evolution
of either, anthropic arguments become plausible only when we cannot
find any other explanation.  That said, I find it difficult to
know whether the cosmological constant problem is severe enough
to justify the explanation of last resort.  

Inflation can conceivably help in the search for a nonanthropic
explanation of vacuum selection, since it offers the possibility
that only a small minority of vacua are populated. Inflation is,
after all, a complicated mechanism that involves exponentially
large factors in its basic description, so it possible that it
populates some states overwhelming more than others.  In
particular, one might expect that those states that lead to the
fastest exponential expansion rates would be favored.  Then these
fastest expanding states --- and their decay products --- could
dominate the multiverse. 

But so far, unfortunately, this is only wishful thinking.  As I
will discuss in the next section, we do not even know how to
define probabilities in eternally inflating multiverses. 
Furthermore, it does not seem likely that any principle that
favors a rapid rate of exponential inflation will favor a vacuum
of the type that we live in.  The key problem, as one might
expect, is the value of the cosmological constant.  The
cosmological constant $\Lambda$ in our universe is extremely
small, i.e., $\Lambda \lta 10^{-120}$ in Planck units.  If
inflation singles out the state with the fastest exponential
expansion rate, the energy density of that state would be
expected to be of order Planck scale or larger.  To explain why
our vacuum has such a small energy density, we would need to find
some reason why this very high energy density state should decay
preferentially to a state with an exceptionally small energy
density \cite{Polchinski-private}. 

There has been some effort to find relaxation methods that might
pick out the vacuum \cite{Steinhardt-Turok2006}, and perhaps this
is the best hope for a nonanthropic explanation of the
cosmological constant.  So far, however, the landscape of
nonanthropic solutions to this problem seems bleak.

\section{Difficulties in Calculating Probabilities}

In an eternally inflating universe, anything that can happen will
happen; in fact, it will happen an infinite number of times. 
Thus, the question of what is possible becomes trivial---anything
is possible, unless it violates some absolute conservation law. 
To extract predictions from the theory, we must therefore learn
to distinguish the probable from the improbable.

However, as soon as one attempts to define probabilities in an
eternally inflating spacetime, one discovers ambiguities.  The
problem is that the sample space is infinite, in that an
eternally inflating universe produces an infinite number of
pocket universes.  The fraction of universes with any particular
property is therefore equal to infinity divided by infinity---a
meaningless ratio.  To obtain a well-defined answer, one needs to
invoke some method of regularization.

To understand the nature of the problem, it is useful to think
about the integers as a model system with an infinite number of
entities.  We can ask, for example, what fraction of the integers
are odd.  Most people would presumably say that the answer is
$1/2$, since the integers alternate between odd and even.  That
is, if the string of integers is truncated after the $N$th, then
the fraction of odd integers in the string is exactly $1/2$ if
$N$ is even, and is $(N+1)/2N$ if $N$ is odd.  In any case, the
fraction approaches $1/2$ as $N$ approaches infinity.

However, the ambiguity of the answer can be seen if one imagines
other orderings for the integers.  One could, if one wished,
order the integers as
\begin{equation}
  1,3,\ 2,\ 5,7,\ 4,\ 9,11,\ 6\ ,\ldots,  
  \label{eq:30}
\end{equation}
always writing two odd integers followed by one even integer. 
This series includes each integer exactly once, just like the
usual sequence ($1,2,3,4, \ldots$).  The integers are just
arranged in an unusual order.  However, if we truncate the
sequence shown in Eq.~(\ref{eq:30}) after the $N$th entry, and
then take the limit $N \to \infty$, we would conclude that 2/3 of
the integers are odd.  Thus, we find that the definition of
probability on an infinite set requires some method of
truncation, and that the answer can depend nontrivially on the
method that is used.

In the case of eternally inflating spacetimes, the natural choice
of truncation might be to order the pocket universes in the
sequence in which they form.  However, we must remember that each
pocket universe fills its own future light cone, so no pocket
universe forms in the future light cone of another.  Any two
pocket universes are spacelike separated from each other, so some
observers will see one as forming first, while other observers
will see the opposite.  One can arbitrarily choose equal-time
surfaces that foliate the spacetime, and then truncate at some
value of $t$, but this recipe is not unique.  In practice,
different ways of choosing equal-time surfaces give different
results. 

\section{The Youngness Paradox}

If one chooses a truncation in the most naive way, one is led to
a set of very peculiar results which I call the {\it youngness
paradox.}

Specifically, suppose that one constructs a Robertson-Walker
coordinate system while the model universe is still in the false
vacuum (de Sitter) phase, before any pocket universes have
formed. One can then propagate this coordinate system forward
with a synchronous gauge condition,\footnote{By a synchronous
gauge condition, I mean that each equal-time hypersurface is
obtained by propagating every point on the previous hypersurface
by a fixed infinitesimal time interval $\Delta t$ in the
direction normal to the hypersurface.} and one can define
probabilities by truncating at a fixed value $t_f$ of the
synchronous time coordinate $t$.  That is, the probability of any
particular property can be taken to be proportional to the volume
on the $t = t_f$ hypersurface which has that property.  This
method of defining probabilities was studied in detail by Linde,
Linde, and Mezhlumian, in a paper with the memorable title ``Do
we live in the center of the world?'' \cite{center-world}.  I
will refer to probabilities defined in this way as synchronous
gauge probabilities.

The youngness paradox is caused by the fact that the volume of
false vacuum is growing exponentially with time with an
extraordinary time constant, in the vicinity of $10^{-37}$ s.
Since the rate at which pocket universes form is proportional to
the volume of false vacuum, this rate is increasing exponentially
with the same time constant.  That means that in each second the
number of pocket universes that exist is multiplied by a factor
of $\exp\left\{10^{37}\right\}$.  At any given time, therefore,
almost all of the pocket universes that exist are universes that
formed very very recently, within the last several time
constants.  The population of pocket universes is therefore an
incredibly youth-dominated society, in which the mature universes
are vastly outnumbered by universes that have just barely begun
to evolve.  Although the mature universes have a larger volume,
this multiplicative factor is of little importance, since in
synchronous coordinates the volume no longer grows exponentially
once the pocket universe forms.

Probability calculations in this youth-dominated ensemble lead to
peculiar results, as discussed in Ref.~\cite{center-world}. 
These authors considered the expected behavior of the mass
density in our vicinity, concluding that we should find ourselves
very near the center of a spherical low-density region.  Here I
would like to discuss a less physical but simpler question, just
to illustrate the paradoxes associated with synchronous gauge
probabilities.  Specifically, I will consider the question: ``Are
there any other civilizations in the visible universe that are
more advanced than ours?''.  Intuitively I would not expect
inflation to make any predictions about this question, but I will
argue that the synchronous gauge probability distribution
strongly implies that there is no civilization in the visible
universe more advanced than us.

Suppose that we have reached some level of advancement, and
suppose that $t_{\rm min}$ represents the minimum amount of time
needed for a civilization as advanced as we are to evolve,
starting from the moment of the decay of the false vacuum---the
start of the big bang.  The reader might object on the grounds
that there are many possible measures of advancement, but I would
respond by inviting the reader to pick any measure she chooses;
the argument that I am about to give should apply to all of them. 
The reader might alternatively claim that there is no sharp
minimum $t_{\rm min}$, but instead we should describe the problem
in terms of a function which gives the probability that, for any
given pocket universe, a civilization as advanced as we are would
develop by time $t$. I believe, however, that the introduction of
such a probability distribution would merely complicate the
argument, without changing the result. So, for simplicity of
discussion, I will assume that there is some sharply defined
minimum time $t_{\rm min}$ required for a civilization as
advanced as ours to develop.

Since we exist, our pocket universe must have an age $t_0$
satisfying
\begin{equation}
  t_0 \ge t_{\rm min} \ . 
  \label{eq:31}
\end{equation}
Suppose, however, that there is some civilization in our pocket
universe that is more advanced than we are, let us say by 1
second.  In that case Eq.~(\ref{eq:31}) is not sufficient, but
instead the age of our pocket universe would have to satisfy
\begin{equation}
  t_0 \ge t_{\rm min} + 1 \hbox{\ second}\ . 
  \label{eq:32}
\end{equation}
However, in the synchronous gauge probability distribution,
universes that satisfy Eq.~(\ref{eq:32}) are outnumbered by
universes that satisfy Eq.~(\ref{eq:31}) by a factor of
approximately $\exp\left\{10^{37}\right\}$.  Thus, if we know
only that we are living in a pocket universe that satisfies
Eq.~(\ref{eq:31}), it is extremely improbable that it also
satisfies Eq.~(\ref{eq:32}).  We would conclude, therefore, that
it is extraordinarily improbable that there is a civilization in
our pocket universe that is at least 1 second more advanced than
we are.

Perhaps this argument explains why SETI has not found any signals
from alien civilizations, but I find it more plausible that it is
merely a symptom that the synchronous gauge probability
distribution is not the right one.

Although the problem of defining probabilities in eternally
inflating universe has not been solved, a great deal of progress
has been made in exploring options and understanding their
properties.  For many years Vilenkin and his collaborators
\cite{vvw, Vilenkin-probabilities} were almost the only
cosmologists working on this issue, but now the field is growing
rapidly \cite{Others-probabilities}.

\section{Does Inflation Need a Beginning?}

If the universe can be eternal into the future, is it possible
that it is also eternal into the past?  Here I will describe a
recent theorem \cite{BGV} which shows, under
plausible assumptions, that the answer to this question is
no.\footnote{There were also earlier theorems about this issue by
Borde and Vilenkin (1994, 1996) \cite{Borde-Vilenkin1,
Borde-Vilenkin2}, and Borde \cite{Borde} (1994), but these
theorems relied on the weak energy condition, which for a perfect
fluid is equivalent to the condition $\rho + p \ge 0$.  This
condition holds classically for forms of matter that are known or
commonly discussed as theoretical proposals.  It can, however, be
violated by quantum fluctuations \cite{Borde-Vilenkin3}, and so
the applicability of these theorems is questionable.}

The theorem is based on the well-known fact that the momentum of
an object traveling on a geodesic through an expanding universe
is redshifted, just as the momentum of a photon is redshifted. 
Suppose, therefore, we consider a timelike or null geodesic
extended backwards, into the past.  In an expanding universe such
a geodesic will be blueshifted.  The theorem shows that under
some circumstances the blueshift reaches infinite rapidity (i.e.,
the speed of light) in a finite amount of proper time (or affine
parameter) along the trajectory, showing that such a trajectory
is (geodesically) incomplete.  

\begin{figure}
   \epsfxsize=240pt
   \centerline{\epsfbox{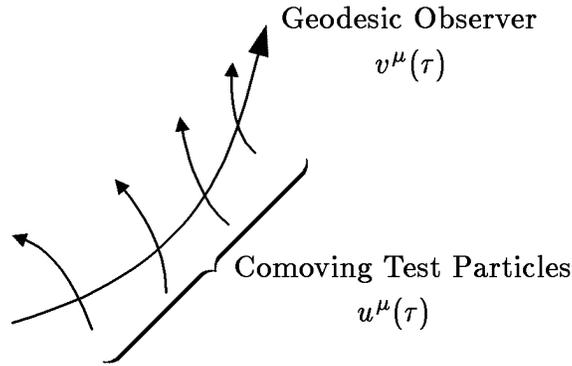}}
   \caption{An observer measures the velocity of passing test
       particles to infer the Hubble parameter.}
   \label{geodesic}
\end{figure}

To describe the theorem in detail, we need to quantify what we
mean by an expanding universe.  We imagine an observer whom we
follow backwards in time along a timelike or null geodesic.  The
goal is to define a local Hubble parameter along this geodesic,
which must be well-defined even if the spacetime is neither
homogeneous nor isotropic.  Call the velocity of the geodesic
observer $v^\mu(\tau)$, where $\tau$ is the proper time in the
case of a timelike observer, or an affine parameter in the case
of a null observer.  (Although we are imagining that we are
following the trajectory backwards in time, $\tau$ is defined to
increase in the future timelike direction, as usual.)  To define
$H$, we must imagine that the vicinity of the observer is filled
with ``comoving test particles,'' so that there is a test
particle velocity $u^\mu(\tau)$ assigned to each point $\tau$
along the geodesic trajectory, as shown in Fig.~\ref{geodesic}. 
These particles need not be real --- all that will be necessary
is that the worldlines can be defined, and that each worldline
should have zero proper acceleration at the instant it intercepts
the geodesic observer. 

To define the Hubble parameter that the observer measures at time
$\tau$, the observer focuses on two particles, one that he passes
at time $\tau$, and one at $\tau + \Delta \tau$, where in the end
he takes the limit $\Delta \tau \rightarrow 0$.  The Hubble
parameter is defined by
\begin{equation}
  H \equiv {\Delta v_{\rm radial} \over \Delta r} \ ,
  \label{eq:39}
\end{equation}
where $\Delta v_{\rm radial}$ is the radial component of the
relative velocity between the two particles, and $\Delta r$ is
their distance, where both quantities are computed in the rest
frame of one of the test particles, not in the rest frame of the
observer.  Note that this definition reduces to the usual one if
it is applied to a homogeneous isotropic universe.

The relative velocity between the observer and the test particles
can be measured by the invariant dot product,
\begin{equation}
  \gamma \equiv u_\mu v^\mu \ ,
  \label{eq:40}
\end{equation}
which for the case of a timelike observer is equal to the usual
special relativity Lorentz factor
\begin{equation}
  \gamma = {1 \over \sqrt{1-v^2_{\rm rel}}} \ .
  \label{eq:41}
\end{equation}

If $H$ is positive we would expect $\gamma$ to decrease with
$\tau$, since we expect the observer's momentum relative to the
test particles to redshift.  It turns out, however, that the
relationship between $H$ and changes in $\gamma$ can be made
precise.  If one defines
\begin{equation}
  F(\gamma) \equiv \cases{1/\gamma & for null observers \cr
      {\rm arctanh}(1/\gamma) & for timelike observers \ ,\cr}
  \label{eq:42}
\end{equation}
then
\begin{equation}
  H = {d F(\gamma) \over d \tau} \ .
  \label{eq:43}
\end{equation}

I like to call $F(\gamma)$ the ``slowness'' of the geodesic
observer, because it increases as the observer slows down,
relative to the test particles.  The slowness decreases as we
follow the geodesic backwards in time, but it is positive
definite, and therefore cannot decrease below zero. 
$F(\gamma)=0$ corresponds to $\gamma = \infty$, or a relative
velocity equal to that of light.  This bound allows us to place a
rigorous limit on the integral of Eq.~(\ref{eq:43}).  For
timelike geodesics,
\begin{equation}
  \int^{\tau_f} \, H \, d \tau \le {\rm arctanh}\left({1 \over
     \gamma_f} \right) = {\rm arctanh}\left( \sqrt{1 - v_{\rm
     rel}^2} \right) \ ,
  \label{eq:44}
\end{equation}
where $\gamma_f$ is the value of $\gamma$ at the final time $\tau
= \tau_f$.  For null observers, if we normalize the affine
parameter $\tau$ by $d \tau / d t = 1$ at the final time
$\tau_f$, then
\begin{equation}
  \int^{\tau_f} \, H \, d \tau \le 1 \ .
  \label{eq:45}
\end{equation}
Thus, if we assume an {\it averaged expansion condition}, i.e.,
that the average value of the Hubble parameter $H_{\rm av}$ along
the geodesic is positive, then the proper length (or affine
length for null trajectories) of the backwards-going geodesic is
bounded.  Thus the region for which $H_{\rm av} > 0$ is
past-incomplete. 

It is difficult to apply this theorem to general inflationary
models, since there is no accepted definition of what exactly
defines this class.  However, in standard eternally inflating
models, the future of any point in the inflating region can be
described by a stochastic model \cite{Goncharov} for inflaton
evolution, valid until the end of inflation.  Except for
extremely rare large quantum fluctuations, $H \gta \sqrt{(8
\pi/3) G \rho_f}$, where $\rho_f$ is the energy density of the
false vacuum driving the inflation.  The past for an arbitrary
model is less certain, but we consider eternal models for which
the past is like the future.  In that case $H$ would be positive
almost everywhere in the past inflating region.  If, however,
$H_{\rm av} > 0$ when averaged over a past-directed geodesic, our
theorem implies that the geodesic is incomplete.

There is of course no conclusion that an eternally inflating
model must have a unique beginning, and no conclusion that there
is an upper bound on the length of all backwards-going geodesics
from a given point.  There may be models with regions of
contraction embedded within the expanding region that could evade
our theorem.  Aguirre and \cite{Aguirre-Gratton2002,
Aguirre-Gratton2003} have proposed a model that evades our
theorem, in which the arrow of time reverses at the $t=-\infty$
hypersurface, so the universe ``expands'' in both halves of the
full de Sitter space.

The theorem does show, however, that an eternally inflating model
of the type usually assumed, which would lead to $H_{\rm av} > 0$
for past-directed geodesics, cannot be complete.  Some new
physics (i.e., not inflation) would be needed to describe the
past boundary of the inflating region.  One possibility would be
some kind of quantum creation event.

One particular application of the theory is the cyclic ekpyrotic
model of Steinhardt \& Turok \cite{Steinhardt-Turok2002}.  This
model has $H_{\rm av} > 0$ for null geodesics for a single cycle,
and since every cycle is identical, $H_{\rm av} > 0$ when
averaged over all cycles.  The cyclic model is therefore
past-incomplete, and requires a boundary condition in the past.

\section{Conclusion}

In this paper I have summarized the arguments that strongly
suggest that our universe is the product of inflation.  I argued
that inflation can explain the size, the Hubble expansion, the
homogeneity, the isotropy, and the flatness of our universe, as
well as the absence of magnetic monopoles, and even the
characteristics of the nonuniformities.  The detailed
observations of the cosmic background radiation anisotropies
continue to fall in line with inflationary expectations, and the
evidence for an accelerating universe fits beautifully with the
inflationary preference for a flat universe.  Our current picture
of the universe seems strange, with 95\% of the energy in forms
of matter that we do not understand, but nonetheless the picture
fits together extraordinarily well.

Next I turned to the question of eternal inflation, claiming that
essentially all inflationary models are eternal. In my opinion
this makes inflation very robust: if it starts anywhere, at any
time in all of eternity, it produces an infinite number of pocket
universes.  A crucial issue in our understanding of fundamental
physics is the selection of the vacuum, which according to
current ideas in string theory could be any one of a colossal
number of possibilities.  Eternal inflation offers at least a
hope that a small set of vacua might be strongly favored.  For
that reason it is important for us to learn more about the
evolution of the multiverse during eternal inflation.  But so far
it is only wishful thinking to suppose that eternal inflation
will allow us to determine the vacuum in which we should expect
to find ourselves.

I then discussed the past of eternally inflating models,
concluding that under mild assumptions the inflating region must
have a past boundary, and that new physics (other than inflation)
is needed to describe what happens at this boundary.  

Although eternal inflation has fascinating consequences, our
understanding of it remains incomplete.  In particular, we still
do not understand how to define probabilities in an eternally
inflating spacetime. 

We should keep in mind, however, that observations in the past
few years have vastly improved our knowledge of the early
universe, and that these new observations have been generally
consistent with the simplest inflationary models.  It is the
success of these predictions that justifies spending time on the
more speculative aspects of inflationary cosmology.

\section*{Acknowledgments}

This work is supported in part by funds provided by the U.S.
Department of Energy (D.O.E.) under grant \#DF-FC02-94ER40818. 
The author would particularly like to thank Joan Sola and his
group at the University of Barcelona, who made the IRGAC-2006
conference so valuable and so enjoyable.


\newif\ifpreprint 
\preprinttrue  
\def\rtitle#1{\ifpreprint ``#1,''\else\ignorespaces\fi}
\def\rtitleNoComma#1{\ifpreprint ``#1''\else\ignorespaces\fi}
\def\arx#1{\ifpreprint [arXiv:#1]\else\unskip\fi}
\def\rtitlep#1{\ifpreprint ``#1,''\else #1\fi}
\def\arxp#1{\ifpreprint arXiv:#1\else {\it Preprint} #1\fi}


\end{document}